\documentclass[11pt,twoside]{article}
\usepackage{FUSE2004}
\usepackage{natbib}

\usepackage{epsf}
\usepackage{psfig}
\usepackage{lscape}
\usepackage{graphicx}

\begin{document}
%\begin{article}

\title{White Dwarfs and Hot Subdwarfs as Seen from FUSE} 

\author{G. Fontaine}
\affil{Universit\'e de Montr\'eal}
\author{P. Chayer}
\affil{The Johns Hopkins University, and University of Victoria}

\begin{abstract}

We present a small collection of FUSE spectra representative of the
main spectral classes found in white dwarf stars. In addition, we
also discuss another family of hot evolved stars, that of the hot 
subdwarfs. Both families belong to the chemically peculiar stars, and it
is thought that a complex interplay of competing processes such as
gravitational settling, ordinary diffusion, radiative levitation, weak
stellar winds, and accretion is responsible for the rich variety of
atmospheric compositions observed in those objects. FUSE is playing a
key role in the current quest for establishing a coherent theory of the
spectral evolution of these stars as it allows the determination of the
patterns of heavy element abundances at a significantly higher level of
accuracy than has been possible before on the basis of optical or UV
observations. We also briefly present some fascinating FUV light curves
of a handful of pulsating subdwarf B stars, thus illustrating the unique
potential of FUSE for asteroseismological studies, a potential which has
not been exploited yet.

\end{abstract}

\section{Introduction}

Despite their intrinsic faintness, white dwarfs and hot subdwarfs are
routinely used as ``lamps'' that provide light along lines of sight in
studies of the ISM, and they are also used during calibration exercises
for many space experiments. This is because their spectra are relatively
simple, owing to the fact that their atmospheres show a high degree of
purity, not seen elsewhere in astrophysics. This is particularly true
for white dwarfs, examples of which can show essentially pure H
atmospheres or pure He atmospheres. This purity is understood as the
result of gravitational settling in the intense gravitational field that
characterizes white dwarfs. Under its influence, all elements heavier
than the dominant atmospheric constituent (either H or He) rapidly sink
below the photosphere and remain out of sight.

However, the record shows that white dwarf atmospheres, although highly
pure, are $not$ perfectly pure in most cases. And indeed, white dwarf
atmospheres are generally polluted by small traces of heavy elements,
particularly at both the high and low effective temperature ends of the
white dwarf cooling sequence. For example, above 30,000 K, it is
generally believed that radiative levitation as well as residual stellar
winds can compete efficiently and $selectively$ against gravitational
settling to produce an amazing variety of atmospheric chemical
compositions. On the other hand, below 20,000 K or so, episodic
fractionated accretion from the ISM is believed to be a temporary source
of heavy metal pollutants that are often detected in the atmospheres of
the cooler white dwarfs.

The competition between gravitational settling, thermal diffusion,
ordinary diffusion, radiative levitation, weak stellar winds,
superficial convection, and accretion produces a bewildering variety of
trace element pollutants along the white dwarf cooling sequence. To a
lesser extent, this is also true of the hot subdwarfs which are also, as
a rule, chemically peculiar stars showing very intriguing atmospheric
compositions. In fact, it could be argued that white dwarfs
and hot subdwarfs constitute the most fascinating of all chemically
peculiar stars. Hence, beside their usefulness as calibrators and lamps for
ISM studies, white dwarfs and hot subdwarfs are certainly of high
intrinsic interest. 

The theory of the spectral evolution of white dwarfs and hot subdwarfs
aims at explaining in a coherent way the puzzling variety of 
surface abundances observed in such stars. This implies detailed
studies of the complex interplay between the various competing
mechanisms that are mentioned just above, mechanisms that are believed
to be at work in those stars. This also implies the establishment of
observed element patterns from which one can make some sense and test
the theory. 

It is here that FUSE is playing its unique role in this particular
endeavor. And indeed, FUSE allows us to detect a full host of traces of
metals in the atmospheres of white dwarfs and hot subdwarfs, traces that
are more often than not invisible in the optical, and that would remain
undetected otherwise. FUSE observations in conjunction with model
atmosphere and spectral synthesis techniques are being used to establish
more firmly than before the patterns of heavy element abundances in
white dwarfs and hot subdwarfs. Already, some very interesting results
have been obtained as reported in these $Proceedings$. Some of these
results challenge our current understanding of the formation of white
dwarf and hot subdwarf spectra. In a broader context, FUSE is also
helping us in refining our detailed understanding of the evolution of
white dwarfs, a necessary step if white dwarf cosmochronology -- i.e.,
the use of white dwarfs as independent age indicators -- is to reach its
potential. 

\section{White Dwarf Stars}

White dwarfs, as is well known, represent the endpoint of the evolution
for the vast majority of stars. It is indeed believed that more than
97\% of the stars in our galaxy (those with initial masses less than
about 8 $M_{\rm \odot}$ on the main sequence) end up as white dwarfs. 
Most of them descend from post-AGB evolution and are former nuclei of
planetary nebulae. They have run out of thermonuclear fuel, and they
consist mainly of C and O, the products of H and He burning. White
dwarfs are thus cooling bodies in hydrostatic equilibrium provided by
the balance between gravity and the pressure of degenerate electrons. 
They shine through the slow leakage of thermal energy coming from the
nondegenerate ions. White dwarfs have a stratified configuration with a
C/O core surrounded by a thin He mantle ($M({\rm He})/M_* \sim
10^{-2}$) itself surrounded by an even thinner H outermost layer
($M({\rm H})/M_* \sim 10^{-4}$). Some 20\% of the white dwarfs, however,
have lost their H layer as they were formed through the so-called
born-again AGB mechanism (a late He flash that destroys the outermost H
layer). Hence, from a spectroscopic point of view, white dwarfs come in
two main ``flavors'', the H-dominated atmosphere (DA) or the
He-dominated atmosphere (non-DA) stars. 

\begin{figure}[!ht]
\plotfiddle{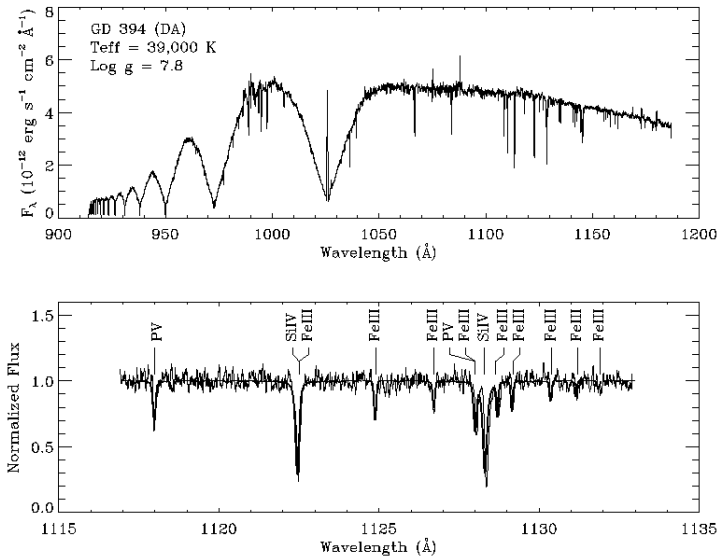}{7cm}{90}{55}{55}{200}{-60}
\caption{Upper panel: FUSE spectrum of the DA white dwarf GD 394 ($V$ =
  13.08). Lower panel: zoom on a spectral region showing the presence of
  heavy element pollutants.}
\end{figure}

A total of about 2200 white dwarfs have been cataloged so far. The
average visual magnitude of those is  $<V>$ $\simeq$ 15.5. They are
found over the full range of effective temperatures in the HR diagram,
from 200,000 K for the hottest and youngest stars, down to 3000 K for the
coolest ones known. They show a narrow mass distribution centered on 0.6 
$M_{\rm \odot}$. This corresponds to a typical radius of about 0.01
$R_{\rm \odot}$, a surface gravity of log $g$ $\sim$ 8, and a mean
density of $<\rho>$ $\sim$ $10^6$ g cm$^{-3}$. The white dwarfs of
spectral type DA are found over almost the full range of effective
temperatures, while the non-DA stars take on the spectral type name
PG1159, DO, DB, DQ, or DC depending on their values of $T_{\rm eff}$.

As an example, Figure 1 shows the FUSE spectrum (upper panel) of GD 394,
a $T_{\rm eff}$ = 39,000 K DA white dwarf. The spectrum is dominated by
the Lyman series of H. The spectral lines are very broad, which is the
signature of the high pressures encountered in the high-gravity
atmospheres characterizing white dwarf stars. The atmosphere of GD 394
is, however, clearly polluted by traces of heavy elements. Indeed, a
zoom on the narrow spectral window shown in the lower panel indicates
the presence of several metals, including P, Si, and Fe. In this
context, it is well to remember here that without competition,
gravitational settling would have left a completely pure H atmosphere in
that star.

Figure 2 shows examples of FUSE spectra for two cooler DA white dwarfs,
one around 30,000 K and the other at 20,000 K. Notice how the width of
the Lyman lines increases considerably with decreasing $T_{\rm eff}$. In
fact, H lines reach their maximum strength at still lower $T_{\rm eff}$
in DA white dwarfs, around 12,000 K. The higher lines of the series become
essentially fused together as can be seen, corresponding physically to
their disappearance in a very high pressure environment. Also, the
spectra shown in Figure 2 show some very interesting features for
experiments in atomic physics. And indeed, the structures identified as
H$_2^+$ correspond to collision-induced quasi-molecular absorption features
whose existence was predicted by Nicole Allard in France. The features
detected by FUSE provide new testing grounds for atomic theory.

\begin{figure}[!ht]
\plotfiddle{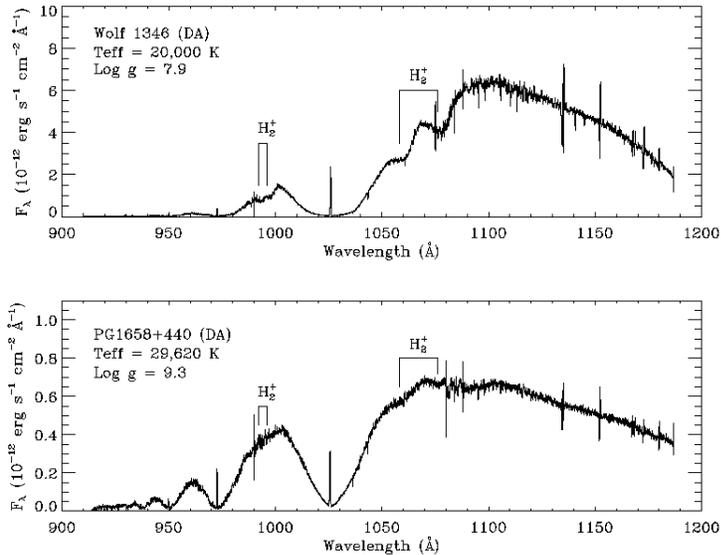}{7cm}{90}{55}{55}{200}{-60}
\caption{Upper panel: FUSE spectrum of the DA white dwarf Wolf 1346 ($V$
  = 11.52). Lower panel: FUSE spectrum of the DA white dwarf PG1658+440
  ($V$ = 14.62).}
\end{figure}

Examples of non-DA spectra are provided in Figure 3. In the top panel,
we show the spectrum of the prototype of the PG1159 class. Note that the
H Lyman lines seen in the spectrum are interstellar in origin here. In
this very hot atmosphere, residual winds prevent settling from operating
efficiently, and the chemical composition is believed to reflect the
composition produced immediately after the born-again AGB phase. It is
mostly a mixture of He, C, and O with traces of several other elements. 
In the paper presented by Klaus Werner in these $Proceedings$, 
one can learn more about these extremely hot and fascinating objects.

In the lower panel of Figure 3, we show an example of a DO spectrum, 
a much cooler star, believed to have descended directly from the PG1159
phase. In the DO phase of the evolution, residual winds have decreased
considerably in intensity, the stars are much older than in the PG1159
phase, and settling is winning its battle. That is, the atmosphere of a
DO star is now dominated by He. The spectrum shows mostly the relatively
strong lines of the Balmer series of HeII.

\begin{figure}[!ht]
\plotfiddle{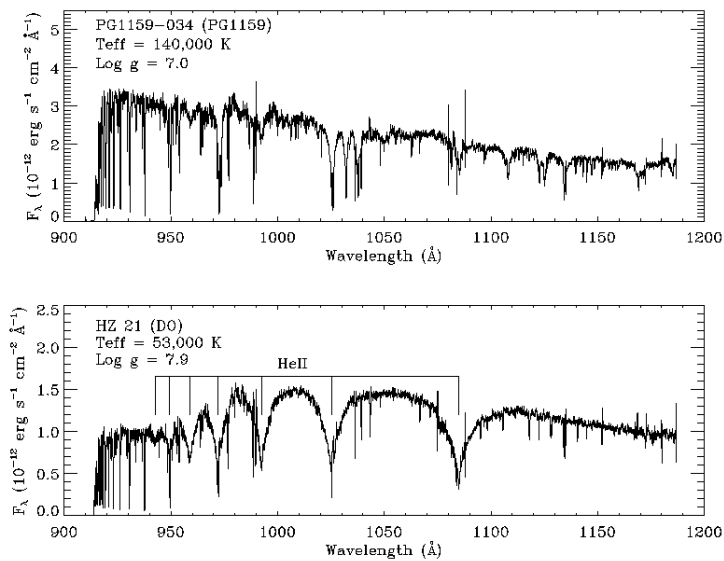}{7cm}{90}{55}{55}{200}{-60}
\caption{Upper panel: FUSE spectrum of the PG1159 white dwarf PG1159$-$034 ($V$
  = 14.87). Lower panel: FUSE spectrum of the DO white dwarf HZ 21 ($V$
  = 14.22 ).}
\end{figure}

\begin{figure}[!ht]
\plotfiddle{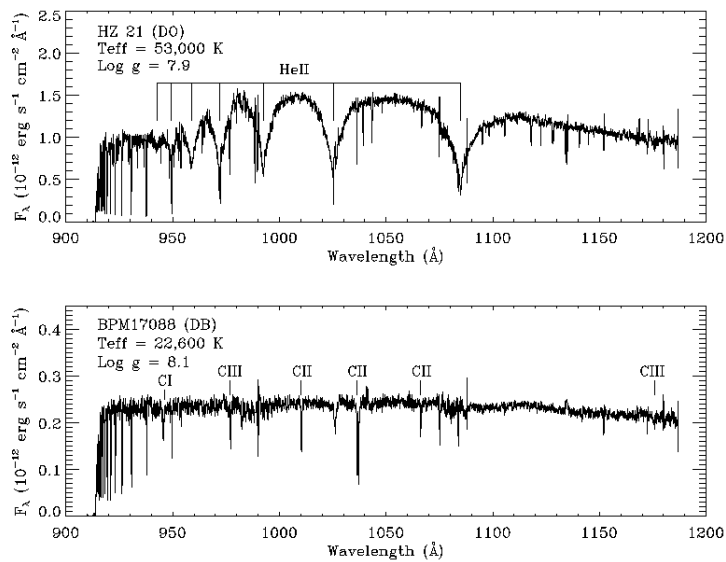}{7cm}{90}{55}{55}{200}{-60}
\caption{Upper panel: FUSE spectrum of the DO white dwarf HZ 21 ($V$ =
  14.22). Lower panel: FUSE spectrum of the DB white dwarf BPM17088 ($V$
  = 14.07).} 
\end{figure}

In the top panel of Figure 4, we again show the spectrum of the DO white
dwarf HZ 21. This is to contrast with its future spectral state in the lower
panel, represented here by the 23,000 K DB star BPM17088. The spectrum of
the latter is rather interesting. Disregarding the interstellar H lines,
it is basically featureless except for some prominent C lines and very little
else! According to the theory of the spectral evolution of white dwarfs,
a DB star should have a perfectly pure He atmosphere, yet this is not
the case here as C is an obvious pollutant. In this respect, we draw 
attention to the interesting contribution by Petitclerc et al. (these
$Proceedings$) where the authors discuss this C pollution problem in
BPM17088 and other hot DB white dwarfs.

\section{Hot Subdwarf Stars}

Most hot subdwarfs belong to spectral type sdB and are evolved, compact
stars that populate the extreme horizontal branch (EHB). They surely
evolved from the RGB, but we still do not know the details of their
formation process, and this constitutes one of the last frontiers of
stellar evolution theory. The sdB stars have atmospheric parameters in 
the range of temperatures from about 20,000 K to upward of 40,000 K, and
in the range of gravities from $\sim$5.0 to $\sim$6.2 in log $g$. They
are believed to be core He-burning objects with masses in a very narrow
range centered around 0.5 $M_{\rm \odot}$. Their residual H-rich
envelopes are too thin to sustain significant H-shell burning during
core He-burning (EHB) evolution. Their envelopes are also too thin to
prevent sdB stars from reaching the AGB after core helium exhaustion,
and they become sdO stars instead ($T_{\rm eff}$ $\sim$ 45,000 K and log $g$
$\sim$ 5.4) in their post-EHB, He shell-burning phase. After a few
$10^8$ yr of evolution near the EHB, hot subdwarfs ultimately collapse
into low-mass white dwarfs, but provide a mere 2\% of the white dwarf
population.  

Hot subdwarf stars are all chemically peculiar. For instance, He is
typically underabundant by 1 to 2 orders of magnitude in sdB stars, and
heavy elements have unusual and puzzling abundances (as inferred from
optical, IUE, EUVE, HST, and FUSE observations). Diffusion processes
(gravitational settling, ordinary diffusion, 
radiative levitation) are believed to be at work in competition with
weak stellar winds. Hot subdwarfs (sdB, sdO) dominate the brighter
end of surveys for faint blue stars down to $V$ $\sim$ 16. They are the
most numerous objects found in shallow surveys for faint blue stars such
as the PG, EC, and MCT surveys (e.g., there are more than 300 hot
subdwarfs brighter than $V$ $\sim$ 14.3 in the PG survey). They are the
main sources of ultraviolet light in old populations such as elliptical
galaxies and some globular clusters (the so-called UV-upturn phenomenon).

One of the current problems in subdwarf B star research is to seek for
an explanation for the observed coexistence of short-period pulsating and
nonpulsating stars in the same region of the log $g$$-$$T_{\rm eff}$
diagram. The variable sdB stars of interest show multiperiodic
luminosity variations with periods in the range 100$-$500 s, which are
caused by pressure-mode pulsational instabilities. Pulsation
calculations have shown that nonradial mode instabilities can be triggered
through the so-called kappa-mechanism associated with a local
overabundance of iron in the envelope of models of pulsating sdB
stars provided the Fe abundance is large enough. Hence, an obvious test of
this idea is to search for a spectral signature for the pulsations, Fe
being the key element.

Along with several collaborators, we have used FUSE to carry out such a
test. It is important to realize that, in order to obtain a meaningful
result, it is necessary to minimize the complicated dependences of
the Fe abundance on log $g$ and $T_{\rm eff}$ in presence of radiative
levitation by comparing the spectrum of a pulsating star with that of a
nonpulsating object with very similar values of their atmospheric
parameters. So far, we have obtained results for one such a pair of
pulsating/nonpulsating stars, and their FUV spectra are shown in Figure 5.

\begin{figure}[!ht]
\plotfiddle{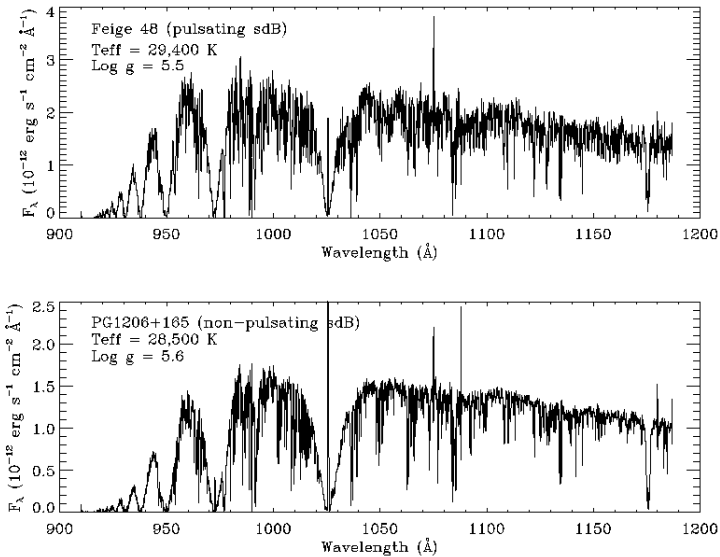}{7cm}{90}{55}{55}{200}{-60}
\caption{Upper panel: FUSE spectrum of the pulsating sdB star Feige 48 ($V$ =
  13.48). Lower panel: FUSE spectrum of the nonpulsating sdB star
  PG1206+165 ($V$ = 13.75). Both stars have similar values of log $g$ and 
  $T_{\rm eff}$, yet one pulsates and the other does not.}
\end{figure}

While the overall spectra of the two objects are quite similar --
consistent with the fact that they have similar values of log $g$ and 
$T_{\rm eff}$ -- it turns out that the pulsating star has a much higher
atmospheric abundance of Fe than the nonpulsating star. This is best
seen in Figures 6 and 7, which show a portion of the spectrum where
prominent Fe lines can be seen. We find, quantitatively, that the Fe
abundance in the pulsating star Feige 48 is more than two orders of
magnitude larger than in its nonpulsating counterpart PG1206+165, in
line with the expectations of pulsation theory. While extremely
encouraging, this result, obtained so far for a single pair of
pulsating/nonpulsating stars, must be confirmed by observing futher
suitable pairs. We hope to be able to do that in the future.

\begin{figure}[!ht]
\plotfiddle{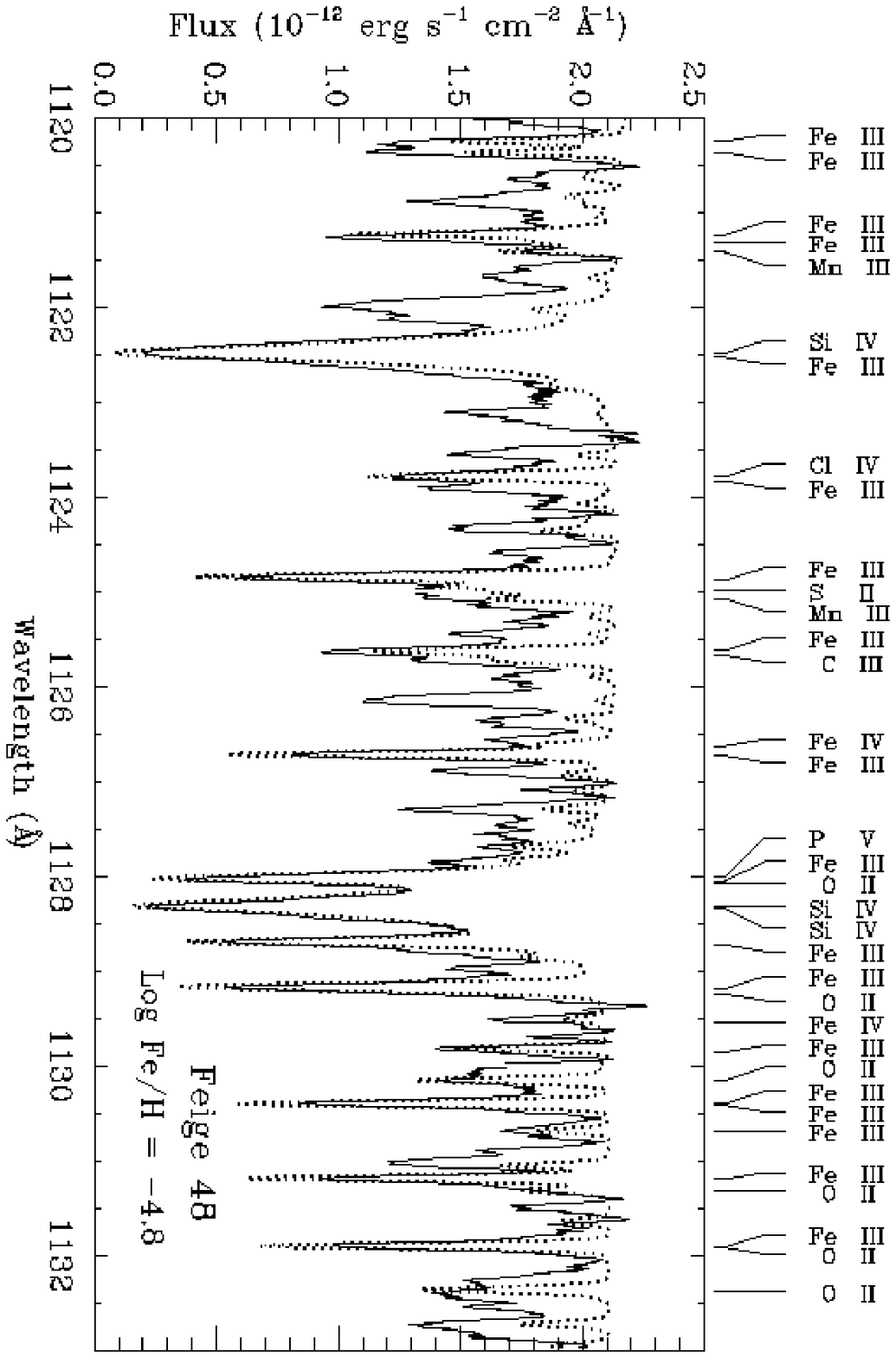}{6cm}{90}{55}{55}{200}{-60}
\caption{Portion of the FUSE spectrum of the pulsating sdB star Feige 48
  showing prominent Fe lines.}
\end{figure}

\begin{figure}[!ht]
\plotfiddle{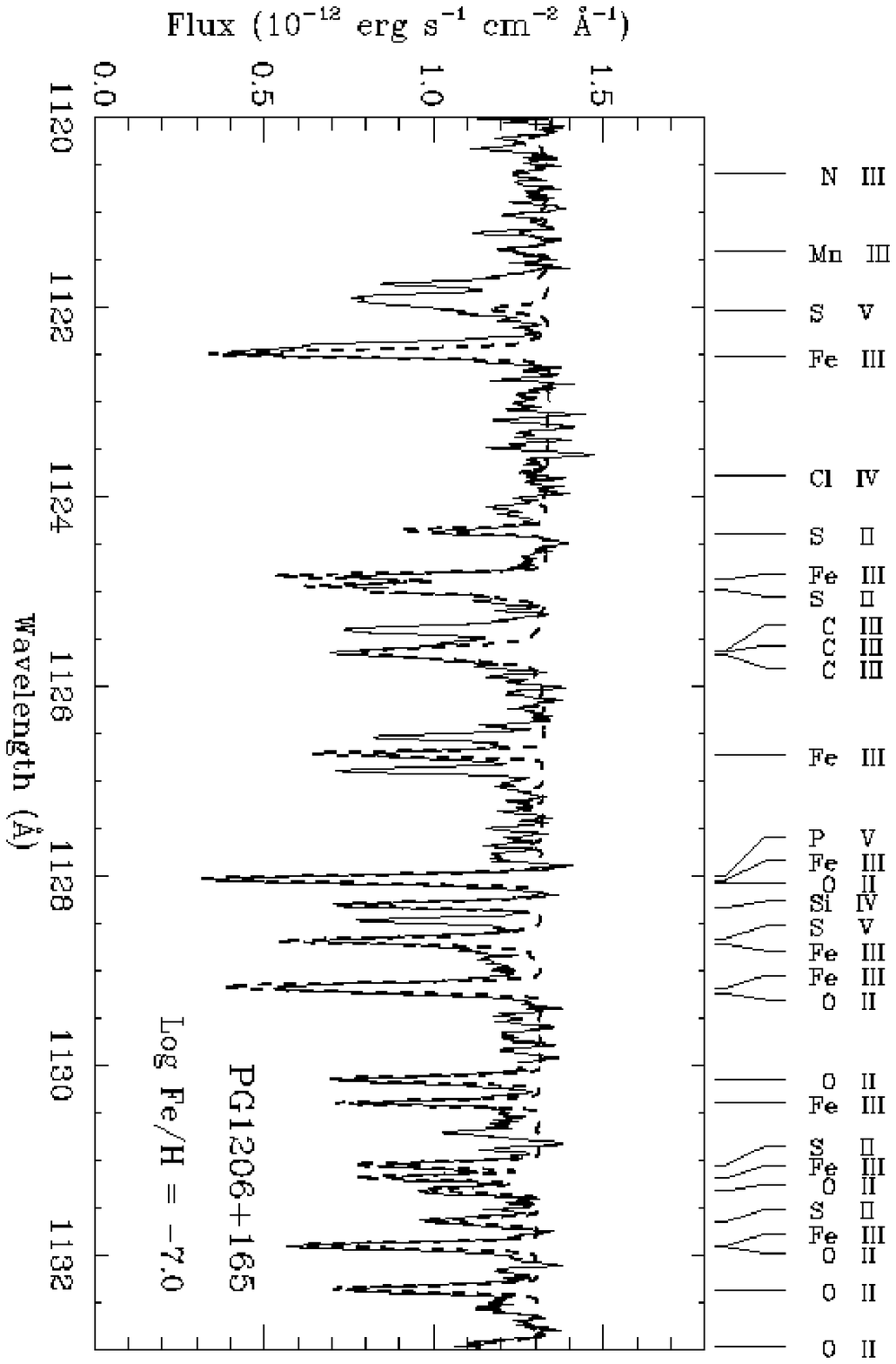}{6cm}{90}{55}{55}{200}{-60}
\caption{Similar to Figure 6, but for the nonpulsating sdB star PG1206+165.}
\end{figure}

\section{FUSE as a FUV Photometer for Asteroseismological Studies}

Our interests in pulsating white dwarfs and hot subdwarfs have led us to
investigate the potential of FUSE as a FUV photometer. One of the key
ingredients in the asteroseismological process is to identify a pulsation
mode in terms of its radial order $k$ (corresponding to the number of
nodes of the eigenfunction in the radial direction) and of its degree
index $\ell$ (corresponding to the azimuthal wave number associated with the
spherical harmonic function which describes the angular geometry of the
mode). The pulsation frequency (period) of a mode depends on these two
``quantum'' numbers for spherically symmetric stellar models. Theory
shows that the ratio of the amplitudes of a given pulsation mode in two
different wavebands is, in a first approximation, independent of the
unknown intrinsic amplitude and of the unknown viewing aspect. This
ratio, however, does depend on the index $\ell$, thus allowing, in
principle, for its empirical determination. For hot stars, calculations 
clearly indicate that the relative mode amplitudes as seen in the
optical and in the FUV strongly discriminates between values of 
the degree $\ell$, allowing these to be easily determined. Such
constraints on the mode identification are otherwise very difficult to
obtain from optical multicolor photometry alone, as the discrimination
between different $\ell$ values is weak when confined to this bandpass.  
The principle is illustrated in Figure 8, as appropriate for a typical
sdB star model with $T_{\rm eff}$ = 30,000 K and $\log g = 
5.5$. The figure shows the relative monochromatic amplitude ratio as a
function of $\ell$ and wavelength for nonradial modes of degree $\ell=1$,
2, 3, and 4.

\begin{figure}[!ht]
\plotfiddle{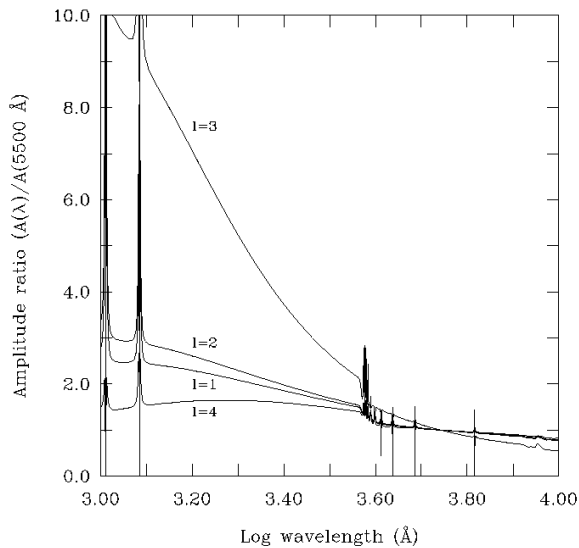}{7.0cm}{0}{50}{50}{-170}{-70}
\caption{Wavelength and $\ell$ dependence of the relative monochromatic
  amplitude (normalized at 5500\AA) of a nonradial pulsation mode of
  degree $\ell=1$, 2, 3, and 4 in a typical sdB model.} 
\end{figure}

With these considerations in mind, we investigated the potential of FUSE
as a FUV photometer for such experiments using data on pulsating sdB
stars obtained previously in program B033 (PI: G. Fontaine). This was
possible as these observations were all gathered in TTAG mode, with a
frequency resolution of 1 Hz. To increase the S/N ratio we integrated
over the full FUSE bandwidth to produce a single FUV ``color'', and we
binned the data in 10 s bins, a sampling time more than adequate for the
periods involved. With the help of B. Godard, we were thus able to
produce the light curves shown in Figure 9. Given the relative faintness
of our target stars and the small aperture of FUSE, these light curves
are, in our view, nothing short of amazing! They establish, above all,
that FUSE can indeed be used as a useful tool for asteroseismological
studies.  

As a comparison, we show, in Figure 10, the corresponding optical light
curves, all obtained in ``white light'' at the 3.6 m CFHT telescope. 
There is already a strong hint that the pulsation amplitudes are larger,
as expected, in the FUV than in the optical domain, but caution should
be taken here because the FUV and optical light curves have not been
taken at the same epoch. Indeed, it is well known that pulsating hot
subdwarfs (and white dwarfs) are multiperiodic pulsators with highly
variable mode amplitudes from season to season. The proper way to use
FUSE as an asteroseismological tool is to gather $contemporaneous$ FUV
and optical observations, and integrate over a sufficient period of time
in order to reach a suitable time resolution in the Fourier domain. The
issue of temporal resolution is essential to be able to isolate
individual frequency peaks (pulsation modes) in that domain. For $each$ of
these modes, the amplitude ratio $A$(FUV)/$A$(optical) would constrain
its $\ell$ value.

\begin{figure}[!ht]
\plotfiddle{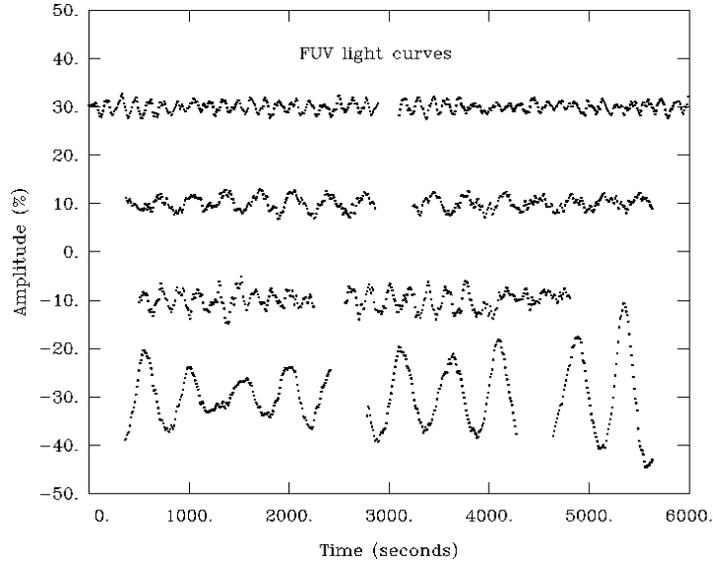}{7.5cm}{-90}{58}{58}{-210}{270}
\caption{FUV light curves of four pulsating sdB stars as obtained by
  FUSE; from top to bottom: PG1219+534 ($V$ = 13.24), Feige 48 ($V$ = 13.48),
  KPD2109+4401 ($V$ = 13.38), and PG1605+072 ($V$ = 12.92). The light curves
  are expressed in terms of percentage of residual amplitude relative to
  the mean brightness of the star. Each plotted point represents a
  sampling time of 10 s. The curves have been shifted by arbitrary
  amounts in the vertical direction away from the zero point for
  visualiztion purposes.}  
\end{figure}

\begin{figure}[!ht]
\plotfiddle{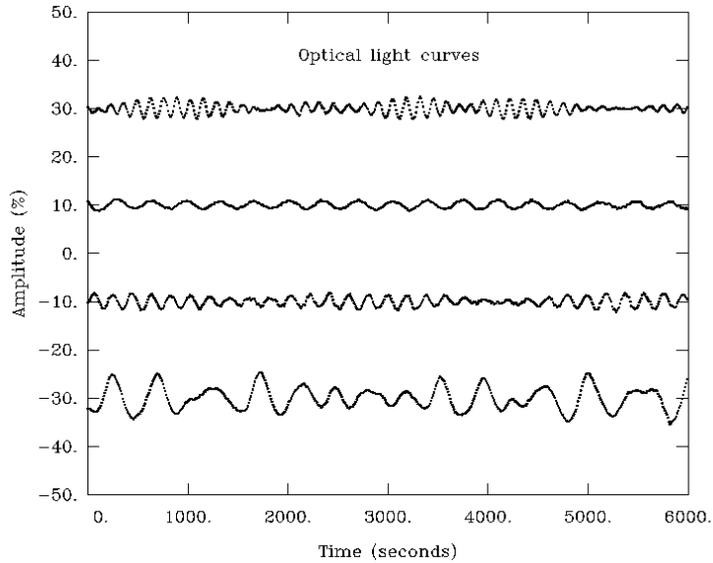}{7.5cm}{-90}{58}{58}{-210}{270}
\caption{Similar to Figure 9, on the same scale and for the same stars,
  but in the optical regime. Here the data are ``white light'' light
  curves gathered at the CFHT with the Montr\'eal 3-channel photometer
  LAPOUNE.} 
\end{figure}

\end{document}